\documentclass[12pt,preprint]{aastex}

\usepackage{graphicx}
\usepackage{txfonts}
\usepackage{longtable}
\usepackage{threeparttable}

\shorttitle{Cosmological time dilation in gamma-ray bursts}
\shortauthors{Zhang et al.}

\begin{document}

\title{Cosmological Time Dilation in Durations of \emph{Swift} Long Gamma-Ray Bursts}

\author{Fu-Wen Zhang\altaffilmark{1,2}, Yi-Zhong Fan\altaffilmark{1}, Lang Shao\altaffilmark{3} and Da-Ming Wei\altaffilmark{1}}
\altaffiltext{1}{Key Laboratory of Dark Matter and Space Astronomy, Purple Mountain Observatory, Chinese Academy of Sciences, Nanjing 210008, China}
\altaffiltext{2}{College of Science, Guilin University of Technology, Guilin 541004, China}
\altaffiltext{3}{Department of Physics, Hebei Normal University, Shijiazhuang 050016, China}

\begin{abstract}

Cosmological time dilation is a fundamental phenomenon in an expanding universe, which stresses that both the duration and wavelength of the emitted light from a distant object at the redshift $z$ will be dilated by a factor of $1+z$ at the observer. By using a sample of 139 \emph{Swift} long GRBs with known redshift ($z\leq8.2$), we measure the observed duration ($T_{90}$) in the observed energy range between $140/(1+z)$ keV and $350/(1+z)$ keV, corresponding to a fixed energy range of 140-350 keV in the rest frame. We obtain a significant correlation between the duration and the factor $1+z$, i.e., $T_{\rm{90}}=10.5(1+z)^{0.94\pm0.26}$, which is well consistent with that expected from cosmological time dilation effect.

\end{abstract}


\section{INTRODUCTION} \label{sec:intro}

Gamma-ray bursts (GRBs) are the most violent explosions in distant galaxies \citep{Piran2004,ZM2004}. The search for the cosmological time dilation signature in GRB data, a fundamental phenomenon in an expanding universe, has a long history. In the pre-{\it Swift}-era, the verification of time dilation signature in GRB data was in heavy debate \citep{Norris1994,Che1997a,Che1997b,Lee1997,Deng1998,Lee2000,Mitrofanov1996,Chang2001,Chang2002,Borgonovo2004,Norris2002,Bloom2003,Wei2003}. The main uncertainty of these early results is that these samples contained small or even zero number of bursts with known redshift. Thanks to the successful performance of {\it Swift} satellite \citep{Gehrels2004}, the number of GRBs with measured redshift increases rapidly and a reliable test of the time dilation signature reported in previous literature becomes possible. Nevertheless, recent analyses still revealed no sound evidence for the cosmological time dilation effect in GRBs detected by {\it Swift} or Fermi Gamma-ray Telescope \citep{Sakamoto2011,Kocevski2013,Gruber2011}.

It is well known that the intrinsic durations or light curves of GRBs are highly energy dependent \citep{Fenimore1995,Norris1996,Peng2006,Zhang2007,Zhang2008,Zhang2012,Qin2013}. We note that most previous works ignored this effect and simply measured the observed durations in a fixed observed energy range. As a result, the received photons belong to different energy ranges when emitted in the rest frame of different GRBs. Therefore, the observed durations would be strongly biased since they simply recorded different parts of the intrinsic light curves. This can be settled by choosing a fixed energy range in the rest frame and measuring the observed duration in a projected energy range by the relation $E_{\rm{obs}}=E_{\rm{rest}}/(1+z)$, where $E_{\rm{obs}}$ and $E_{\rm{rest}}$ are the energy of the photon measured in the observer and the rest frame, respectively \citep{Sakamoto2011,Gruber2011,Ukwatta2012}. When taking this effect into account, we calculate the observed durations of \emph{Swift} GRBs with known redshifts within the observed energy band $140/(1+z)$ keV to $350/(1+z)$ keV, corresponding to the same rest frame energy range 140-350 keV, and reanalyze the redshift dependence of the durations. We find that there is a significant trend that the inferred duration tend to be longer in bursts at higher redshifts and the durations are  stretched approximately by a factor $(1+z)$, as expected from cosmological time dilation effect. We describe our sample and data analysis in section 2, present the results in Section 3, and give our conclusions in Section 4.

\section{Sample and data analysis} \label{sec:analysis}

In order to obtain a complete sample and minimize the influence of different instruments (with different sensitivities and energy bands), only \emph{Swift} GRBs with known redshift are considered. We obtain a sample of 194 bursts with known redshift\footnote{http://www.mpe.mpg.de/$\sim$jcg/grbgen.html} detected by March, 2012. We download the data from \emph{Swift} Archive available at ftp://legacy.gsfc.nasa.gov/swift/data/. The time tagged event (TTE) data from Burst Alert Telescope (BAT) onboard \emph{Swift} have excellent time resolution of 100 $\mu$s, which can be used to perform the temporal analysis well. The standard BAT software (HEASOFT 6.8) and the latest calibration database are used to process the BAT TTE data. We extract 64 ms (long bursts) or 16 ms (short bursts) binned light curve from the TTE data and determine the GRB duration, $T_{90}$ ($T_{50}$), by the time in which $90\%$ ($50\%$) of the burst counts are collected \citep{Kouveliotou1993}. The widely used Bayesian Block method \citep{Scargle1998} is adopted to extract the duration value.

Using the standard method described above, we firstly calculate the values of $T_{\rm{90,raw}}$ and $T_{\rm{50,raw}}$, where the subscript ``raw" represents the data measured in the observer's energy range of $15-350$ keV. These duration values are generally used in previous studies. But the statistical analysis of $T_{\rm 90,raw}$ and $T_{\rm 50,raw}$ is somehow meaningless or even misleading, because the values of $T_{\rm 90,raw}$ and $T_{\rm 50,raw}$ are highly affected by both the energy-dependent effect and the cosmological time-dilation effect. By fixing the energy range in the GRB rest frame, the energy-dependent effect is removed. We then created the light curves in the observed energy range $140/(1+z)$ keV to $350/(1+z)$ keV. This energy band was chosen so that the projecting energy bands of all {\it Swift} GRBs with known redshifts lie in the {\it Swift}-BAT observed energy range (15-350 keV). We used the same algorithm to find the best $T_{90}$ and $T_{50}$ durations in the observed $140/(1+z)$ - $350/(1+z)$ keV band. The rest frame durations $T_{\rm 90,rest}$ and $T_{\rm 50,rest}$ can be easily obtained by dividing the observed durations by $(1+z)$.

In our initial sample of 194 GRBs, some bursts are not bright enough to measure $T_{\rm{90}}$ and $T_{50}$ in the rest frame energy range $140-350$ keV. Six super-long/peculiar bursts (GRBs 060124, 060218, 100316D, 101225A, 110328A, 111209A) are excluded in our investigation. In the analysis we also exclude the short duration bursts, including two with extended emission (GRBs 060614 and 061210). Please note the short and long duration bursts ($\lessgtr$2 s) are defined following Kouveliotou et al. \citep{Kouveliotou1993}\footnote{Recently, Bromberg et al. (2013) suggested that the commonly used limit of 2 s is conservative and the division $\lessgtr$0.8 is more suitable for $Swift$.}, not by the $T_{90}$ measured in this work. The benefit of excluding them is to have a sample of GRBs that have an intrinsically same (or similar) duration distribution. Our analysis is thus based on a {\it Swift} GRB sample consisting of 139 long GRBs. In addition, three very high redshift candidates (GRBs 090429B, 120521C and 120923A) are also presented for comparison.

\section{Relation between duration and redshift} \label{subsec:dur}

\begin{figure}
\begin{center}
\includegraphics[width=150mm,angle=0]{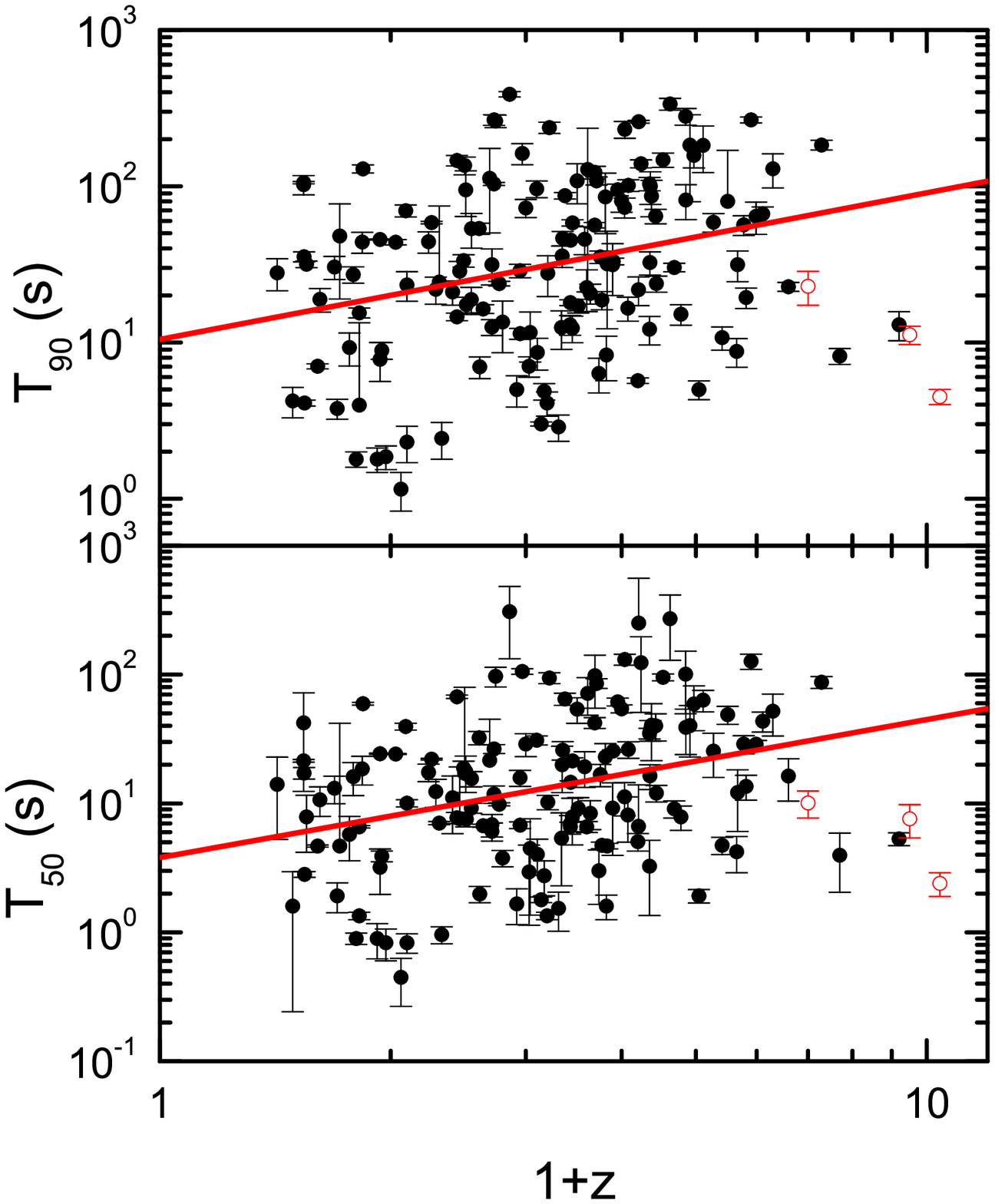}
\end{center}
\caption{Correlations between the durations ($T_{\rm{90}}$ and $T_{\rm{50}}$ measured in the observed energy range of $140/(1+z) -350/(1+z)$ keV) and redshift. The solid lines are our best fits: in the upper panel, we have $T_{\rm{90}}=10.5(1+z)^{0.94}$ with $r=0.29$ and $p=0.0005$; in the lower panel, we have $T_{\rm{50}}=3.8(1+z)^{1.07}$ with $r=0.32$ and $p=0.0001$. Three high redshift candidates (GRBs 090429B, 120521C, and 120923A) are also presented (open circles).} \label{fig:1}
\end{figure}

Assuming that the intrinsic duration of all GRBs is similar, one would expect the observed duration to increase as a function of redshift due to the cosmological expansion. As shown in Fig.\ref{fig:1}, there is a clear trend that farther bursts tend to have larger $T_{90}$ and $T_{50}$. We parameterize the correlation and obtain log$ T_{\rm{90}}=(1.02\pm0.14)+(0.94\pm0.26)$ log$(1+z)$, where the Pearson correlation coefficient is $r=0.29$ and the chance probability is $p=0.0005$. For $T_{\rm{50}}$, we have log$T_{\rm{50}}=(0.58\pm0.14)+(1.07\pm0.27)$log$(1+z)$ with $r=0.32$ and $p=0.0001$. Therefore the observed GRB durations are indeed stretched approximately by a factor of $(1+z)$, as expected from the cosmological time dilation effect.
The scatter is large and particularly two very high redshift GRBs (GRBs 080913A at $z=6.7$ and 090423 at $z=8.2$) and two high redshift candidates (GRBs 090429B at $z\sim9.4$ and 120923A at $z\sim8.5$) do not comply with the correlation well. This might be because the intrinsic duration is not the same for all bursts. Besides, Zhang et al. (2009) have showed that most GRBs with the highest redshifts seem to have rest-frame durations shorter than 2 s, yet still show multi-wavelength properties similar to most long GRBs. Recently, by the simulations, several groups found that the diminishing S/N of higher redshift GRBs makes only the bright narrow portions of the bursts accessible to the detectors (i.e., the so-called ``tip-of-iceberg'' effect), so the measured durations should be considered as lower limits to the true values (Kocevski \& Petrosian 2013; Lv et al. 2012; Littlejohns et al. 2013).

\begin{figure}
\begin{center}
\includegraphics[width=150mm,angle=0]{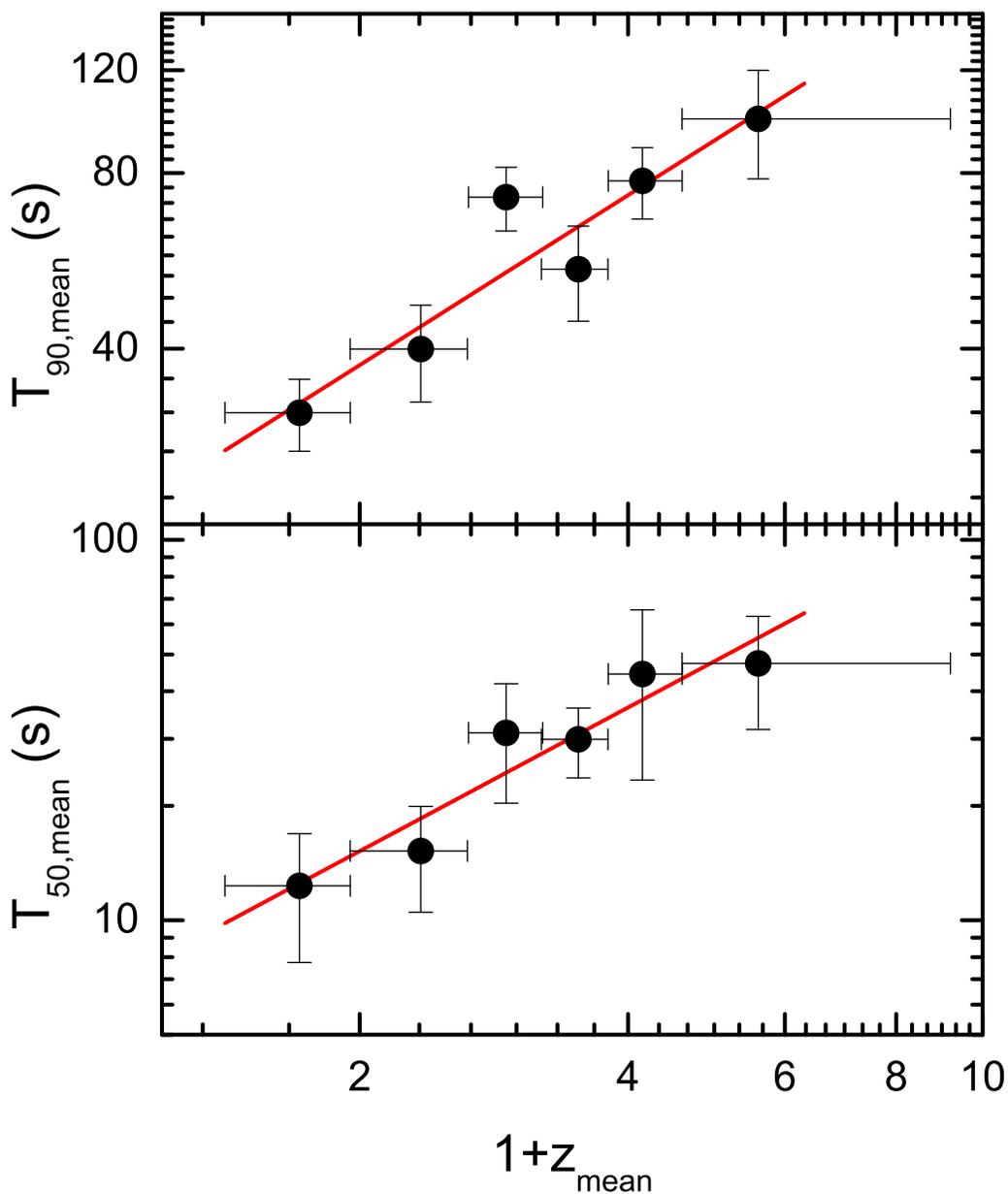}
\end{center}
\caption{Correlations between the mean durations ($T_{\rm{90,mean}}$ and $T_{\rm{50,mean}}$) and the mean redshift ($z_{\rm mean}$) for six groups with almost the same number of bursts, where the errors of $z_{\rm mean}$ only represent the redshift range in each group. The solid lines are our best fits: $T_{\rm{90,mean}}=19.1(1+z_{\rm mean})^{0.97}$ with $r=0.93$ and $p=0.007$ (upper panel) and $T_{\rm{50,mean}}=6.3(1+z_{\rm mean})^{1.25}$ with $r=0.95$ and $p=0.004$ (lower panel).} \label{fig:2}
\end{figure}

To better show the correlation we divide the sample of 139 GRBs into six groups with almost equal number of bursts. We calculate the mean values of $T_{\rm{90}}$, $T_{\rm{50}}$ and $z$ in each groups and reanalyze their relations. From Fig.\ref{fig:2} we find that the mean durations ($T_{\rm{90,mean}}$, $T_{\rm{50,mean}}$) are tightly correlated with the mean redshift $z_{\rm mean}$. Fitting the correlation we have log$ T_{\rm{90,mean}}=(1.28\pm0.10)+(0.97\pm0.19)$ log$(1+z_{\rm mean})$ with $r=0.93$ and $p=0.007$, and log$T_{\rm{50,mean}}=(0.80\pm0.11)+(1.25\pm0.20)$log$(1+z_{\rm mean})$ with $r=0.95$ and $p=0.004$. Hence the $(1+z)$ stretching of durations is established. We have also analyzed the potential influence of the numbers of groups on the statistical result and found that the slope of the correlation is almost invariable and close to 1 for the different sets of groups. Although the intrinsic durations of individual bursts are very different, their mean value is dilated exactly by a factor of 1+z following the nature of the expanding universe \citep{Paczynski1992,Piran1992}.

\begin{figure}
\begin{center}
\includegraphics[width=150mm,angle=0]{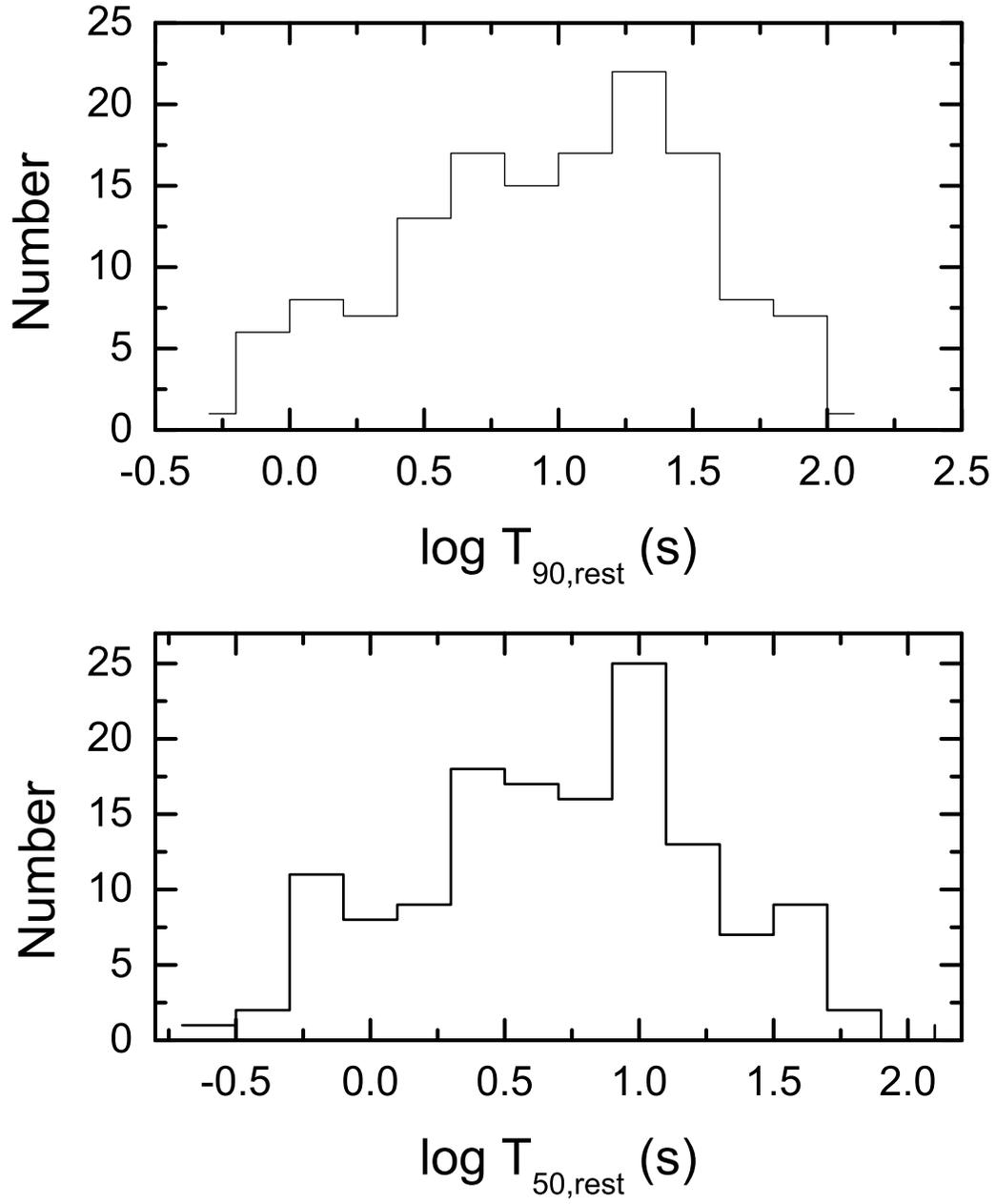}
\end{center}
\caption{Distributions of the rest frame durations $T_{\rm{90,rest}}$ and $T_{\rm{50,rest}}$.}
\label{fig:3}
\end{figure}

\begin{figure}
\begin{center}
\includegraphics[width=150mm,angle=0]{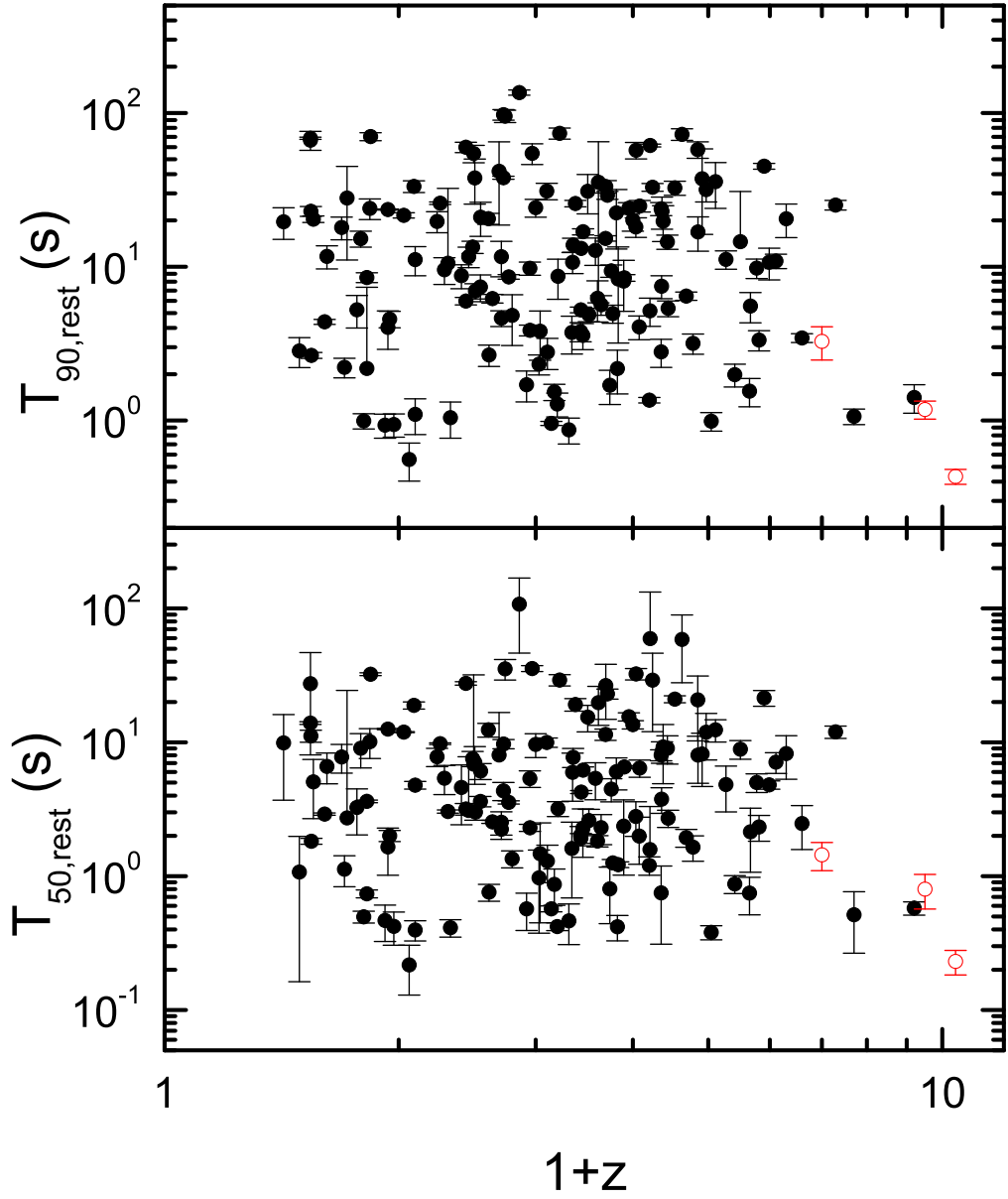}
\end{center}
\caption{Relations between $T_{\rm{90,rest}}$, $T_{\rm{50,rest}}$ and redshift. Other symbols are the same as Fig. 1.}
\label{fig:4}
\end{figure}

Is the observed duration stretching due to the redshift evolution of the intrinsic duration of GRBs? To answer this question we also analyze the distribution of the rest frame duration $T_{\rm 90,rest}$ and $T_{\rm 50,rest}$ as well as the relation between these two quantities and redshift. From Fig.\ref{fig:3}, we find that the distributions of $T_{\rm 90,rest}$ and $T_{\rm 50,rest}$ all span a wide range and their log-median values are 10.7 s and 4.6 s, respectively. Obviously, different GRBs do not have a standard intrinsic duration. But, the similar median values of the intrinsic duration $T_{\rm 90,rest}\sim 10$ s have been reported by many authors even though different energy ranges and different instruments are engaged \citep{Pelangeon2008,Shao2010,Gruber2011}. Therefore, we can only identify the cosmological time dilation as a statistical effect. Fig.\ref{fig:4} shows the redshift dependence of $T_{\rm 90,rest}$ and $T_{\rm 50,rest}$ and we do not find any evidence of the evolution effect of the rest frame duration, where the correlation coefficients (chance probabilities) between $T_{\rm{90,rest}}$ and $T_{\rm{50,rest}}$ and the redshifts are $r=-0.02$ (p= 0.82) and $r=0.02$ ($p=0.79$), respectively. A similar conclusion was also obtained from analyzing the preliminary Fermi/GBM data \citep{Gruber2011}.

It is well known that the duration of GRBs is highly affected by the detector threshold. In order to avoid the influence of the detector threshold on the correlation between duration and redshift, we construct a subsample with relatively bright bursts in the 15-150 keV Swift/BAT band. The subsample is selected with the criteria that the bursts have 1-s peak photon flux $P\geq2.6$ ph s$^{-1}$ cm$^{-2}$ as done by Salvaterra et al.(2012). 63 GRBs match our selection criteria. We then analyze the relations between $T_{90}$, $T_{50}$ and $z$ (Fig. 5). From Fig. 5, we can find that the observed durations are highly depend on redshift for these bright GRBs, which is consistent with the above result. We parameterize the correlation and obtain log$ T_{\rm{90}}=(0.87\pm0.20)+(1.07\pm0.45)$ log$(1+z)$ with $r=0.29$ and $p=0.02$. For $T_{\rm{50}}$, we have log$T_{\rm{50}}=(0.46\pm0.19)+(1.01\pm0.42)$log$(1+z)$ with $r=0.29$ and $p=0.02$. These results further confirm that the cosmological time dilation effect identified in the duration of GRBs is reliable.

Previous works in analyzing the data of \emph{Swift} GRBs have reported no evidence for the duration being stretched by a factor of $(1+z)$ \citep{Sakamoto2011,Kocevski2013}, which seems to be at odds with our results. To check what happens, following previous approaches we also investigate the correlations between $T_{\rm{90,raw}}$, $T_{\rm{50,raw}}$ and $z$ (Fig. 6). From Fig. 6, we find that there is indeed no evidence for the dilation-like effect in the raw duration data, in agreement with that found in previous studies. The respective correlation coefficients between $T_{\rm{90,raw}}$ and $T_{\rm{50,raw}}$ and the redshifts are $r=0.03$ (p= 0.69) and $r=0.13$ ($p=0.13$). This suggests that the cosmological time dilation effect has been canceled by the energy-dependent effect of the duration, since the farther the burst is located, the shorter portion of the light curve (corresponding to a higher energy range in the rest frame) would be recorded in the observed energy range.

In addition, P{\'e}langeon et al. (2008) and Kocevski \& Petrosian (2013) found that the rest frame duration decrease as a function of redshift (see also Wei \& Gao 2003). However, it should be noted that the rest frame duration used by these works only simply dividing the raw duration measured in a fixed detector energy range by a factor of $(1+z)$, the energy-dependent effect is not considered. As a test, we also calculate $T_{\rm{90,raw}}/(1+z)$ and $T_{\rm{50,raw}}/(1+z)$ and analyze their relations with redshift. As shown in Fig. 7, both $T_{\rm{90,raw}}/(1+z)$ and $T_{\rm{50,raw}}/(1+z)$ all show a decreasing trend with increasing redshift. We parameterize the correlations and obtain log$(T_{\rm{90,raw}}/(1+z))=(1.65\pm0.13)+(-0.92\pm0.25)$ log$(1+z)$ with $r=-0.3$ and $p=0.0004$, and log$(T_{\rm{50,raw}}/(1+z))=(1.07\pm0.14)+(-0.64\pm0.27)$log$(1+z)$ with $r=-0.2$ and $p=0.02$. Hence we have demonstrated that a reliable relation between the duration and the redshift can not be reliably established if one ignores the energy-dependent effect.

\section{Summary and conclusions}

In this work we perform a statistical analysis of the duration of a sample of 139 long GRBs with known redshift detected by \emph{Swift} until March, 2012. We calculated the observed duration ($T_{90}$ and $T_{50}$) of all bursts in the observed energy range $140/(1+z)$ keV to $350/(1+z)$ keV, which correspond to a fixed energy bands 140-350 keV in the rest frame. This actually means that the energy-dependent effect is removed. By analyzing the relation between $T_{90}$, $T_{50}$ and redshift, we find that there is a significant trend that both $T_{\rm{90}}$ and $T_{\rm{50}}$ tend to be longer in bursts at higher redshifts and $T_{\rm{90}}=10.5(1+z)^{0.94\pm0.26}$ and $T_{\rm{50}}=3.8(1+z)^{1.07\pm0.27}$. Such results are well consistent with that expected from cosmological time dilation effect that all timescales of GRBs should be stretched by a factor of $(1+z)$. We also find that the intrinsic duration of GRBs is independent with redshift and its distribution span a wide range, where the median value of $T_{\rm 90,rest}$ ($T_{\rm 50,rest}$) is 10.7 s (4.6 s), respectively. If one only uses the raw duration calculated within a fixed detector energy range to make the statistical analysis of duration, the result can be misleading. For example in some literature the ``intrinsic duration" is found to be anti-correlated with the redshift, which is at odds with our finding. Hence a reliable relation between the duration and the redshift can not be reliably established if one ignores the energy-dependent effect of the duration.

We note that the correlation between duration and redshift has a very large scatter, this might be due to the intrinsic scatter in duration. A more important fact is that the several very high redshift GRBs deviate from the correlation. This might be caused by the integrated effect. A more important reason is the "tip-of-iceberg" effect that with increasing redshift and decreasing signal-to-noise ratio only the brightest portion of GRB light curves can be detected, so the measured durations should be considered as lower limits to the true values (Kocevski \& Petrosian 2013; Lv et al. 2012; Littlejohns et al. 2013). In addition, the BAT effective area is not uniform, it sharply drops above 100 keV, and below 25 keV (Barthelmy et al. 2005; see also http://swift.gsfc.nasa.gov/analysis/bat\_digest.html). This mainly affect the two extremes of the redshift distribution. The durations of low-redshift (z$<$1) and highest-redshift (z$>$8) events could therefore be also underestimated.

\acknowledgments

We thank the anonymous referee for the insightful comments/suggestions. This work was supported in part by the National Basic Research Program of China (No. 2014CB845800 and No. 2013CB837000) and the National Natural Science Foundation of China (grants 11163003, U1331101, 11073057, 11103083 and 11273063). Y.-Z.F. is also supported by the 100 Talents program of the Chinese Academy of Sciences and by the Foundation for Distinguished Young Scholars of Jiangsu Province, China (No. BK2012047). F.-W.Z. also acknowledges the support by the China Postdoctoral Science Foundation funded project (No.~20110490139), the Guangxi Natural Science Foundation (No.~2013GXNSFAA019002) and the doctoral research foundation of Guilin University of Technology.

\clearpage



\begin{figure}
\begin{center}
\includegraphics[width=130mm,angle=0]{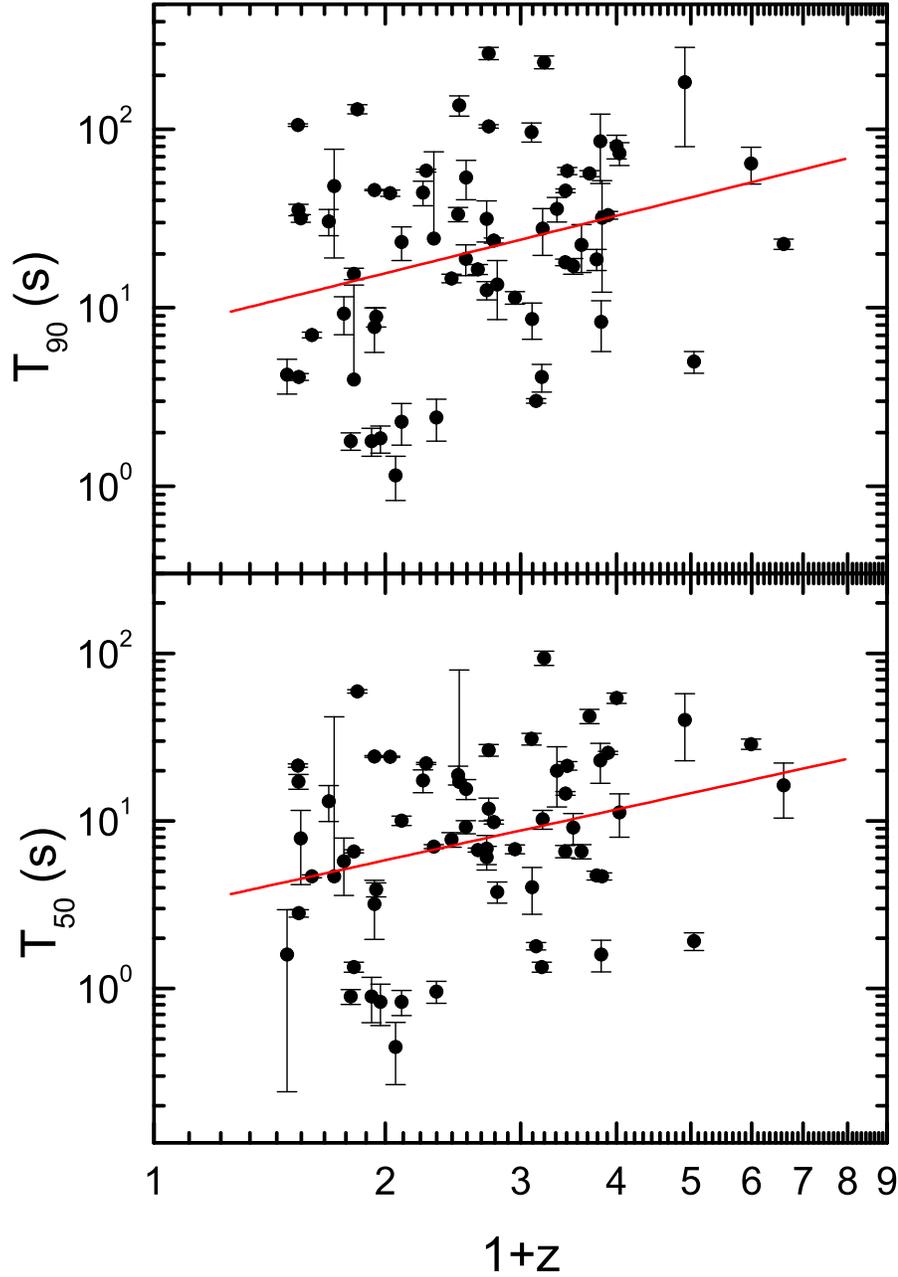}
\end{center}
\caption{Correlations between the durations ($T_{\rm{90}}$ and $T_{\rm{50}}$) and redshifts for 63 relatively bright GRBs with 1-s peak photon flux $P\geq2.6$ ph s$^{-1}$ cm$^{-2}$ in the 15-150 keV energy bands. The solid lines are our best fits: in the upper panel, we have $T_{\rm{90}}=7.4(1+z)^{1.07}$ with $r=0.29$ and $p=0.02$; in the lower panel, we have $T_{\rm{50}}=2.9(1+z)^{1.01}$ with $r=0.29$ and $p=0.02$. } \label{fig:5}
\end{figure}

\begin{figure}
\begin{center}
\includegraphics[width=150mm,angle=0]{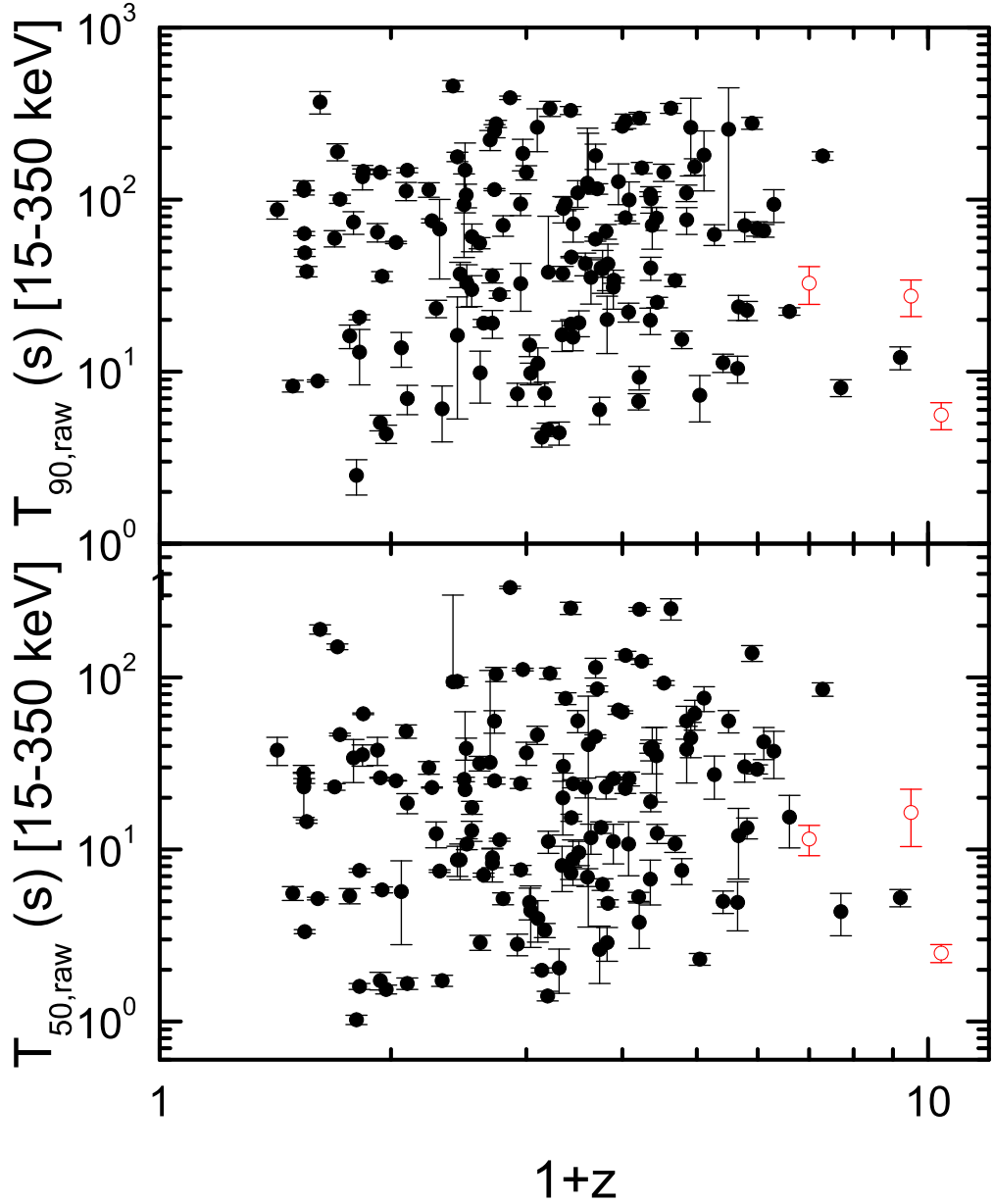}
\end{center}
\caption{Relations between the raw duration ($T_{\rm{90,raw}}$ and $T_{\rm{50,raw}}$) and redshift for 139 BAT long GRBs with known redshift. The values of $T_{\rm{90,raw}}$ and $T_{\rm{50,raw}}$ are calculated directly in the BAT detector energy range 15-350 keV. Other symbols are the same as Fig. 1.} \label{fig:6}
\end{figure}

\begin{figure}
\begin{center}
\includegraphics[width=150mm,angle=0]{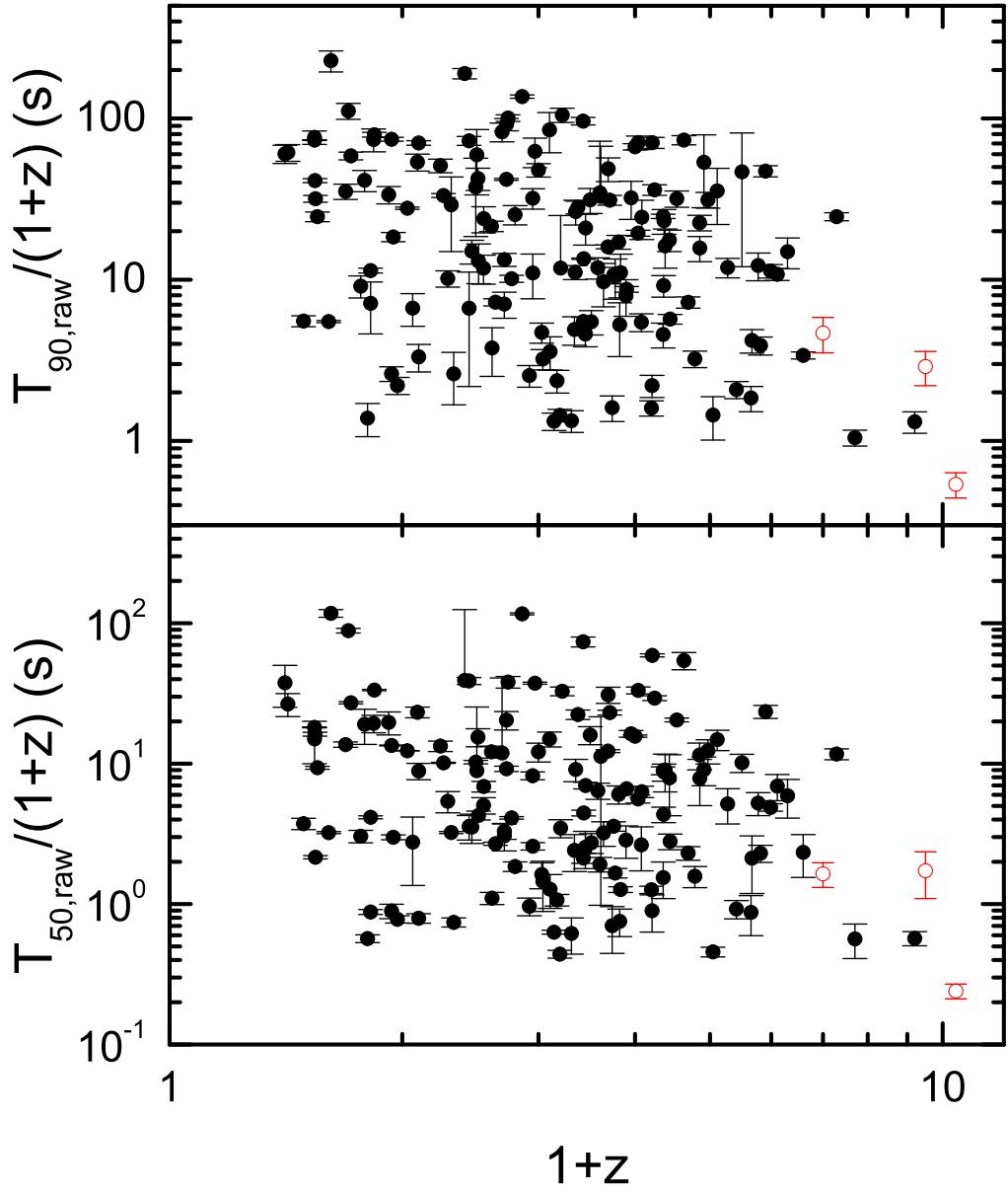}
\end{center}
\caption{Correlations between the raw durations ($T_{\rm{90,raw}}/(1+z)$ and $T_{\rm{50,raw}}/(1+z)$ )in the rest frame and redshift. Other symbols are the same as Fig. 1.} \label{fig:7}
\end{figure}

\clearpage

\begin{deluxetable}{lllllllll}
\tabletypesize{\scriptsize}
\tablecaption{Durations of 142 Swift long GRBs with known redshift$^{*}$}
\tablewidth{0pt}
\tablehead{
\colhead{GRB} &
\colhead{Obs ID} &
\colhead{z} &
\colhead{$T_{90,raw}$} &
\colhead{$T_{50,raw}$} &
\colhead{$E_{1}^{\star}$} &
\colhead{$E_{2}^{\star}$} &
\colhead{$T_{90}$} &
\colhead{$T_{50}$} \\
\colhead{name} &
\colhead{} &
\colhead{} &
\colhead{(s)} &
\colhead{(s)} &
\colhead{keV} &
\colhead{keV} &
\colhead{(s)}&
\colhead{(s)}
}
\startdata

    050126 &     103780 &       1.29 &      23.3$\pm$2.7  &      12.4$\pm$2.1  &         61 &        153 &      21.9$\pm$4.3  &      12.4$\pm$3.0  \\

    050315 &     111063 &      1.949 &      94.5$\pm$13.7  &      24.2$\pm$1.5  &         47 &        119 &      28.8$\pm$2.8  &      15.8$\pm$2.2  \\

    050318 &     111529 &       1.44 &      16.3$\pm$10.9  &       8.7$\pm$1.7  &         57 &        143 &      14.6$\pm$0.8  &       7.7$\pm$0.8  \\

    050319 &     111622 &       3.24 &     153.0$\pm$11.4  &     124.2$\pm$4.9  &         33 &         83 &     139.4$\pm$8.2  &     123.6$\pm$72.7  \\

    050401 &     113120 &        2.9 &      34.0$\pm$1.4  &      25.9$\pm$0.5  &         36 &         90 &      33.0$\pm$1.6  &      25.5$\pm$0.6  \\

    050505 &     117504 &       4.27 &      62.8$\pm$8.7  &      27.3$\pm$7.7  &         27 &         66 &      58.8$\pm$8.0  &      25.5$\pm$9.6  \\

   050525A &     130088 &      0.606 &       8.8$\pm$0.1  &       5.2$\pm$0.1  &         87 &        218 &       7.0$\pm$0.3  &       4.7$\pm$0.1  \\

    050603 &     131560 &      2.821 &      20.1$\pm$7.4  &       2.9$\pm$0.6  &         37 &         92 &       8.3$\pm$2.6  &       1.6$\pm$0.3  \\

    050730 &     148225 &       3.97 &     155.5$\pm$17.3  &      61.5$\pm$12.1  &         28 &         70 &     157.6$\pm$26.6  &      59.2$\pm$22.5  \\

    050802 &     148646 &       1.71 &      19.1$\pm$3.5  &       8.3$\pm$1.8  &         52 &        129 &      12.5$\pm$1.5  &       6.1$\pm$1.0  \\

    050803 &     148833 &      0.422 &      87.5$\pm$10.4  &      37.8$\pm$7.0  &         98 &        246 &      27.9$\pm$6.5  &      14.1$\pm$8.8  \\

    050814 &     150314 &        5.3 &      94.0$\pm$20.3  &      37.2$\pm$11.4  &         22 &         56 &     129.3$\pm$31.9  &      52.0$\pm$18.6  \\

   050820A &     151207 &      2.612 &     118.6$\pm$124.5  &      40.7$\pm$37.2  &         39 &         97 &     128.0$\pm$ 106.9  &      71.5$\pm$23.1  \\

    050904 &     153514 &       6.29 &     179.5$\pm$10.3  &      85.4$\pm$7.6  &         19 &         48 &     183.6$\pm$13.2  &      87.2$\pm$9.2  \\

    050908 &     154112 &       3.35 &      19.9$\pm$3.5  &       6.7$\pm$2.0  &         32 &         80 &      12.2$\pm$2.5  &       3.3$\pm$1.9  \\

   050922C &     156467 &      2.198 &       4.6$\pm$0.4  &       1.4$\pm$0.1  &         44 &        109 &       4.1$\pm$0.7  &       1.3$\pm$0.1  \\

   051109A &     163136 &      2.346 &      37.2$\pm$3.7  &      21.2$\pm$8.4  &         42 &        105 &      35.8$\pm$5.6  &      20.0$\pm$7.8  \\

    051111 &     163438 &      1.549 &      61.0$\pm$11.2  &      17.5$\pm$1.5  &         55 &        137 &      53.6$\pm$13.4  &      15.6$\pm$2.1  \\

    060108 &     176453 &       2.03 &      14.3$\pm$2.0  &       4.9$\pm$1.0  &         46 &        116 &       7.0$\pm$1.0  &       2.9$\pm$1.8  \\

    060115 &     177408 &       3.53 &     144.1$\pm$16.6  &      92.8$\pm$2.9  &         31 &         77 &     147.5$\pm$15.4  &      95.2$\pm$5.7  \\

    060206 &     180455 &      4.045 &       7.3$\pm$2.2  &       2.3$\pm$0.2  &         28 &         69 &       5.0$\pm$0.7  &       1.9$\pm$0.2  \\

    060210 &     180977 &       3.91 &     263.0$\pm$125.6  &      44.5$\pm$10.3  &         29 &         71 &     183.3$\pm$ 103.6  &      40.2$\pm$17.3  \\

   060223A &     192059 &       4.41 &      11.3$\pm$1.4  &       5.0$\pm$0.7  &         26 &         65 &      10.8$\pm$1.8  &       4.7$\pm$0.7  \\

    060418 &     205851 &      1.489 &      93.6$\pm$47.5  &      25.5$\pm$0.7  &         56 &        141 &      33.4$\pm$3.1  &      18.8$\pm$2.4  \\

   060502A &     208169 &       1.51 &      32.8$\pm$9.1  &      10.8$\pm$0.8  &         56 &        139 &      17.5$\pm$2.0  &       7.6$\pm$1.0  \\

   060510B &     209352 &        4.9 &     278.3$\pm$22.1  &     138.6$\pm$14.7  &         24 &         59 &     265.9$\pm$11.7  &     126.7$\pm$17.1  \\

    060522 &     211117 &       5.11 &      66.0$\pm$5.6  &      42.2$\pm$9.0  &         23 &         57 &      66.6$\pm$7.1  &      43.5$\pm$7.7  \\

    060526 &     211957 &       3.21 &     297.9$\pm$23.6  &     248.8$\pm$6.3  &         33 &         83 &     258.8$\pm$5.4  &     250.8$\pm$308.0  \\

   060602A &     213180 &      0.787 &      73.9$\pm$11.1  &      34.0$\pm$9.6  &         78 &        196 &      27.2$\pm$3.2  &      16.1$\pm$4.6  \\

    060605 &     213630 &       3.78 &      15.4$\pm$1.8  &       7.6$\pm$1.3  &         29 &         73 &      15.2$\pm$2.3  &       7.9$\pm$1.7  \\

   060607A &     213823 &      3.082 &      99.6$\pm$27.2  &      25.8$\pm$2.4  &         34 &         86 &     101.3$\pm$8.7  &      26.2$\pm$3.9  \\

    060707 &     217704 &       3.43 &      78.1$\pm$11.9  &      35.1$\pm$16.2  &         32 &         79 &      64.3$\pm$6.8  &      40.0$\pm$9.5  \\

    060708 &     217805 &       1.92 &       7.4$\pm$1.1  &       2.8$\pm$0.4  &         48 &        120 &       5.0$\pm$1.1  &       1.7$\pm$0.5  \\

    060714 &     219101 &       2.71 &     115.6$\pm$8.1  &      85.8$\pm$3.5  &         38 &         94 &     108.2$\pm$6.4  &      85.4$\pm$7.4  \\

    060729 &     221755 &       0.54 &     117.6$\pm$11.2  &      23.1$\pm$7.7  &         91 &        227 &     102.5$\pm$14.4  &      42.2$\pm$29.8  \\

    060814 &     224552 &       0.84 &     146.0$\pm$4.9  &      61.5$\pm$0.7  &         76 &        190 &     129.3$\pm$7.7  &      59.4$\pm$1.5  \\

   060904B &     228006 &      0.703 &     189.7$\pm$21.5  &     150.8$\pm$5.7  &         82 &        206 &       3.8$\pm$0.6  &       1.9$\pm$0.5  \\

    060906 &     228316 &      3.685 &      33.9$\pm$2.7  &      10.8$\pm$1.2  &         30 &         75 &      30.2$\pm$1.9  &       9.1$\pm$1.4  \\

    060908 &     228581 &       2.43 &      18.9$\pm$1.2  &       7.3$\pm$0.6  &         41 &        102 &      18.0$\pm$0.8  &       6.6$\pm$0.6  \\

   060912A &     229185 &      0.937 &       5.1$\pm$0.5  &       1.7$\pm$0.2  &         72 &        181 &       7.8$\pm$2.2  &       3.2$\pm$1.2  \\

    060926 &     231231 &      3.208 &       9.3$\pm$1.5  &       3.8$\pm$1.1  &         33 &         83 &      21.8$\pm$4.5  &       6.7$\pm$7.3  \\

    060927 &     231362 &        5.6 &      22.4$\pm$1.1  &      15.4$\pm$5.2  &         21 &         53 &      22.7$\pm$1.5  &      16.3$\pm$5.9  \\

    061007 &     232683 &      1.261 &      75.3$\pm$2.1  &      22.9$\pm$0.2  &         62 &        155 &      58.6$\pm$1.1  &      22.1$\pm$0.3  \\

   061110B &     238174 &       3.44 &      25.3$\pm$1.8  &      12.4$\pm$1.6  &         32 &         79 &      23.8$\pm$2.8  &      12.0$\pm$1.8  \\

    061121 &     239899 &      1.314 &      67.5$\pm$32.9  &       7.5$\pm$0.1  &         61 &        151 &      24.4$\pm$50.4  &       7.0$\pm$0.2  \\

   061222B &     252593 &      3.355 &      40.1$\pm$6.1  &      18.9$\pm$2.4  &         32 &         80 &      32.5$\pm$5.5  &      16.4$\pm$3.6  \\

    070110 &     255445 &      2.352 &      88.7$\pm$15.5  &      30.5$\pm$5.5  &         42 &        104 &      46.5$\pm$4.9  &      25.9$\pm$4.2  \\

    070306 &     263361 &      1.497 &     148.5$\pm$64.7  &      22.3$\pm$41.0  &         56 &        140 &     136.1$\pm$17.7  &      17.2$\pm$62.5  \\

    070318 &     271019 &      0.836 &     136.0$\pm$22.0  &      35.5$\pm$5.0  &         76 &        191 &      44.0$\pm$6.7  &      18.6$\pm$4.7  \\

    070411 &     275087 &      2.954 &     127.4$\pm$33.9  &      64.6$\pm$3.6  &         35 &         89 &      95.0$\pm$8.2  &      61.4$\pm$4.3  \\

    070506 &     278693 &       2.31 &       4.4$\pm$0.7  &       2.0$\pm$0.6  &         42 &        106 &       2.9$\pm$0.6  &       1.5$\pm$0.5  \\

    070508 &     278854 &       0.82 &      20.7$\pm$0.7  &       7.6$\pm$0.2  &         77 &        192 &      15.5$\pm$1.1  &       6.6$\pm$0.1  \\

    070521 &     279935 &      0.553 &      38.2$\pm$2.7  &      14.5$\pm$0.3  &         90 &        225 &      31.7$\pm$1.5  &       7.9$\pm$3.7  \\

    070529 &     280706 &     2.4996 &     109.4$\pm$19.4  &      55.9$\pm$8.2  &         40 &        100 &     108.3$\pm$31.1  &      54.0$\pm$12.1  \\

    070611 &     282003 &       2.04 &       9.8$\pm$1.4  &       4.4$\pm$1.7  &         46 &        115 &      11.6$\pm$4.1  &       4.5$\pm$3.1  \\

   070612A &     282066 &      0.617 &     369.4$\pm$55.0  &     190.3$\pm$11.8  &         87 &        216 &      18.9$\pm$3.2  &      10.7$\pm$2.8  \\

   070714B &     284856 &       0.92 &      64.6$\pm$7.9  &      37.8$\pm$7.0  &         73 &        182 &       1.8$\pm$0.3  &       0.9$\pm$0.3  \\

   070721B &     285654 &      3.626 &     339.9$\pm$22.2  &     251.0$\pm$35.5  &         30 &         76 &     336.0$\pm$29.1  &     271.4$\pm$142.6  \\

    070802 &     286809 &       2.45 &      15.9$\pm$2.6  &       8.8$\pm$2.6  &         41 &        101 &      12.4$\pm$2.4  &       7.9$\pm$3.1  \\

   070810A &     287364 &       2.17 &       7.5$\pm$1.2  &       3.4$\pm$0.3  &         44 &        110 &       4.9$\pm$0.6  &       2.8$\pm$0.8  \\

    071003 &     292934 &        1.1 &     148.1$\pm$4.7  &      18.6$\pm$2.5  &         67 &        167 &      23.4$\pm$5.0  &      10.0$\pm$0.7  \\

   071010B &     293795 &      0.947 &      35.8$\pm$2.3  &       5.8$\pm$0.2  &         72 &        180 &       8.9$\pm$1.1  &       3.9$\pm$0.4  \\

    071020 &     294835 &      2.142 &       4.2$\pm$0.5  &       2.0$\pm$0.1  &         45 &        111 &       3.0$\pm$0.1  &       1.8$\pm$0.1  \\

    071031 &     295670 &      2.692 &     180.1$\pm$30.2  &     114.2$\pm$15.0  &         38 &         95 &     122.8$\pm$11.4  &      98.0$\pm$43.2  \\

    071117 &     296805 &      1.331 &       6.1$\pm$2.2  &       1.7$\pm$0.1  &         60 &        150 &       2.4$\pm$0.6  &       1.0$\pm$0.1  \\

    080207 &     302728 &       1.74 &     275.6$\pm$13.1  &     104.5$\pm$9.9  &         51 &        128 &     261.5$\pm$24.2  &      96.9$\pm$16.9  \\

    080210 &     302888 &      2.641 &      35.3$\pm$10.6  &      11.7$\pm$2.3  &         38 &         96 &      20.6$\pm$3.0  &       8.4$\pm$2.1  \\

    080310 &     305288 &     2.4266 &     330.2$\pm$17.6  &     252.9$\pm$20.5  &         41 &        102 &      13.0$\pm$1.9  &       6.9$\pm$1.1  \\

   080319B &     306757 &      0.937 &     144.0$\pm$4.0  &      26.1$\pm$0.2  &         72 &        181 &      45.6$\pm$0.4  &      24.3$\pm$0.3  \\

   080319C &     306778 &       1.95 &      32.5$\pm$10.0  &       7.6$\pm$0.5  &         47 &        119 &      11.4$\pm$0.9  &       6.8$\pm$0.4  \\

    080411 &     309010 &       1.03 &      56.4$\pm$0.9  &      25.1$\pm$0.1  &         69 &        172 &      43.8$\pm$1.7  &      24.2$\pm$0.2  \\

   080413A &     309096 &      2.433 &      46.3$\pm$0.4  &      15.3$\pm$0.7  &         41 &        102 &      45.2$\pm$1.1  &      14.6$\pm$0.4  \\

   080413B &     309111 &        1.1 &       7.0$\pm$1.4  &       1.7$\pm$0.1  &         67 &        167 &       2.3$\pm$0.6  &       0.8$\pm$0.1  \\

    080430 &     310613 &      0.767 &      16.1$\pm$2.5  &       5.4$\pm$0.5  &         79 &        198 &       9.3$\pm$2.2  &       5.8$\pm$2.2  \\

    080516 &     311762 &        3.2 &       6.7$\pm$0.7  &       5.3$\pm$0.3  &         33 &         83 &       5.7$\pm$0.2  &       5.1$\pm$0.8  \\

   080603B &     313087 &       2.69 &      59.0$\pm$1.7  &      45.4$\pm$1.1  &         38 &         95 &      56.4$\pm$2.3  &      42.2$\pm$4.1  \\

    080605 &     313299 &     1.6398 &      19.1$\pm$1.0  &       7.1$\pm$0.2  &         53 &        133 &      16.4$\pm$1.0  &       6.7$\pm$0.2  \\

    080607 &     313417 &      3.036 &      78.5$\pm$3.3  &      22.7$\pm$1.5  &         35 &         87 &      73.3$\pm$10.7  &      11.3$\pm$3.3  \\

    080721 &     317508 &      2.602 &     124.6$\pm$135.8  &       6.9$\pm$0.8  &         39 &         97 &      22.5$\pm$6.7  &       6.6$\pm$0.6  \\

    080804 &     319016 &     2.2045 &      37.8$\pm$42.3  &      11.1$\pm$1.6  &         44 &        109 &      27.8$\pm$8.1  &      10.2$\pm$1.3  \\

    080805 &     319036 &      1.505 &     106.3$\pm$16.7  &      38.7$\pm$5.7  &         56 &        140 &      94.8$\pm$30.8  &      17.7$\pm$5.6  \\

    080810 &     319584 &       3.35 &     107.6$\pm$3.3  &      38.8$\pm$2.5  &         32 &         80 &     103.7$\pm$10.1  &      34.9$\pm$3.3  \\

   080905B &     323898 &      2.374 &      94.9$\pm$8.6  &      75.5$\pm$5.9  &         41 &        104 &      87.0$\pm$4.2  &      64.6$\pm$6.9  \\

    080906 &     323984 &          2 &     143.7$\pm$13.7  &      36.4$\pm$5.6  &         47 &        117 &      72.6$\pm$9.7  &      28.9$\pm$5.8  \\

    080913 &     324561 &        6.7 &       8.1$\pm$0.9  &       4.4$\pm$1.2  &         18 &         45 &       8.2$\pm$1.0  &       4.0$\pm$1.9  \\

   080916A &     324895 &      0.689 &      59.6$\pm$6.5  &      23.1$\pm$1.0  &         83 &        207 &      30.4$\pm$5.1  &      13.1$\pm$3.2  \\

    080928 &     326115 &      1.692 &     222.5$\pm$29.7  &      32.1$\pm$77.8  &         52 &        130 &     112.5$\pm$62.2  &      21.6$\pm$23.3  \\

    081008 &     331093 &     1.9685 &     185.7$\pm$38.6  &     111.1$\pm$1.9  &         47 &        118 &     162.2$\pm$25.0  &     105.7$\pm$5.3  \\

   081028A &     332851 &      3.038 &     286.5$\pm$27.6  &     134.3$\pm$7.6  &         35 &         87 &     231.2$\pm$28.3  &     130.9$\pm$12.8  \\

    081029 &     332931 &     3.8479 &     109.4$\pm$12.1  &      55.9$\pm$12.1  &         29 &         72 &     280.3$\pm$34.3  &     100.7$\pm$51.1  \\

    081118 &     334877 &       2.58 &      42.5$\pm$6.4  &      23.0$\pm$3.0  &         39 &         98 &      45.8$\pm$9.2  &      19.3$\pm$5.9  \\

    081121 &     335105 &      2.512 &      19.3$\pm$3.3  &       9.6$\pm$1.5  &         40 &        100 &      17.2$\pm$1.7  &       9.2$\pm$1.9  \\

   081203A &     336489 &        2.1 &     263.6$\pm$73.6  &      46.4$\pm$5.6  &         45 &        113 &      96.3$\pm$11.8  &      30.9$\pm$2.5  \\

    081222 &     337914 &       2.77 &      39.6$\pm$11.1  &       6.3$\pm$0.5  &         37 &         93 &      18.7$\pm$2.5  &       4.7$\pm$0.3  \\

    090102 &     338895 &      1.547 &      30.0$\pm$6.1  &      12.9$\pm$1.7  &         55 &        137 &      18.8$\pm$3.7  &       9.2$\pm$0.8  \\

    090205 &     342121 &     4.6497 &      10.4$\pm$1.9  &       4.9$\pm$1.6  &         25 &         62 &       8.8$\pm$1.8  &       4.2$\pm$1.3  \\

    090313 &     346386 &      3.375 &      70.9$\pm$19.4  &      39.4$\pm$11.9  &         32 &         80 &      86.5$\pm$22.7  &      40.4$\pm$19.0  \\

   090418A &     349510 &      1.608 &      56.0$\pm$4.2  &      31.7$\pm$3.1  &         54 &        134 &      53.6$\pm$4.4  &      32.3$\pm$3.8  \\

    090423 &     350184 &        8.2 &      12.1$\pm$1.8  &       5.2$\pm$0.6  &         15 &         38 &      13.0$\pm$2.7  &       5.3$\pm$0.6  \\

    090424 &     350311 &      0.544 &      49.1$\pm$2.4  &       3.3$\pm$0.1  &         91 &        227 &       4.1$\pm$0.2  &       2.8$\pm$0.1  \\

   090516A &     352190 &      4.109 &     181.6$\pm$69.3  &      75.8$\pm$12.5  &         27 &         69 &     182.8$\pm$60.3  &      63.4$\pm$12.0  \\

    090519 &     352648 &       3.85 &      76.3$\pm$13.6  &      38.2$\pm$13.8  &         29 &         72 &      81.7$\pm$20.5  &      38.9$\pm$15.0  \\

    090618 &     355083 &       0.54 &     113.1$\pm$0.6  &      27.9$\pm$0.2  &         91 &        227 &     105.5$\pm$1.7  &      21.4$\pm$0.5  \\

   090715B &     357512 &          3 &     267.0$\pm$12.0  &      62.8$\pm$1.1  &         35 &         88 &      80.3$\pm$12.1  &      54.2$\pm$3.9  \\

    090809 &     359530 &      2.737 &       6.0$\pm$1.1  &       2.6$\pm$1.0  &         37 &         94 &       6.3$\pm$1.6  &       3.0$\pm$1.4  \\

    090812 &     359711 &      2.452 &      72.3$\pm$15.6  &      24.1$\pm$1.3  &         41 &        101 &      58.3$\pm$2.7  &      21.4$\pm$1.2  \\

   090926B &     370791 &       1.24 &     114.2$\pm$11.1  &      29.9$\pm$2.5  &         63 &        156 &      44.2$\pm$6.9  &      17.5$\pm$2.7  \\

    091018 &     373172 &      0.971 &       4.4$\pm$0.5  &       1.5$\pm$0.1  &         71 &        178 &       1.9$\pm$0.3  &       0.8$\pm$0.2  \\

    091020 &     373458 &       1.71 &      36.1$\pm$3.3  &       9.0$\pm$1.1  &         52 &        129 &      31.5$\pm$8.2  &       6.8$\pm$1.4  \\

    091024 &     373674 &      1.092 &     112.3$\pm$13.6  &      48.6$\pm$4.3  &         67 &        167 &      69.7$\pm$6.1  &      39.5$\pm$2.4  \\

    091029 &     374210 &      2.752 &      40.1$\pm$4.7  &      13.4$\pm$1.0  &         37 &         93 &      35.3$\pm$3.2  &      16.8$\pm$3.3  \\

   091109A &     375246 &      3.076 &      22.2$\pm$2.8  &      10.8$\pm$3.7  &         34 &         86 &      16.6$\pm$2.9  &       8.1$\pm$3.1  \\

    091127 &     377179 &       0.49 &       8.3$\pm$0.6  &       5.6$\pm$0.5  &         94 &        235 &       4.2$\pm$0.9  &       1.6$\pm$1.4  \\

   091208B &     378559 &      1.063 &      13.8$\pm$3.2  &       5.7$\pm$2.9  &         68 &        170 &       1.2$\pm$0.3  &       0.4$\pm$0.2  \\

   100219A &     412982 &     4.6667 &      23.8$\pm$4.0  &      12.0$\pm$5.3  &         25 &         62 &      31.5$\pm$7.0  &      12.2$\pm$6.1  \\

   100302A &     414592 &      4.813 &      22.7$\pm$2.9  &      13.4$\pm$1.9  &         24 &         60 &      19.4$\pm$2.9  &      13.6$\pm$3.0  \\

   100513A &     421814 &      4.772 &      70.7$\pm$13.8  &      30.3$\pm$5.6  &         24 &         61 &      56.6$\pm$8.2  &      28.9$\pm$4.6  \\

   100621A &     425151 &      0.542 &      63.5$\pm$1.8  &      25.0$\pm$0.8  &         91 &        227 &      35.4$\pm$2.6  &      17.2$\pm$1.8  \\

   100728B &     430172 &      2.106 &      11.1$\pm$2.6  &       4.0$\pm$1.1  &         45 &        113 &       8.6$\pm$2.0  &       4.0$\pm$1.3  \\

   100814A &     431605 &       1.44 &     177.4$\pm$11.2  &      94.8$\pm$5.4  &         57 &        143 &     146.1$\pm$11.2  &      67.3$\pm$2.0  \\

   100816A &     431764 &     0.8034 &       2.5$\pm$0.6  &       1.0$\pm$0.1  &         78 &        194 &       1.8$\pm$0.2  &       0.9$\pm$0.1  \\

   100901A &     433065 &      1.408 &     457.7$\pm$33.9  &      94.3$\pm$206.7  &         58 &        145 &      21.1$\pm$3.7  &      11.1$\pm$5.2  \\

   100902A &     433160 &        4.5 &     256.6$\pm$190.4  &      55.9$\pm$8.3  &         25 &         64 &      80.1$\pm$89.6  &      48.9$\pm$7.8  \\

   100906A &     433509 &      1.727 &     114.3$\pm$1.6  &      55.7$\pm$8.3  &         51 &        128 &     103.4$\pm$2.3  &      11.8$\pm$1.8  \\

   110128A &     443861 &      2.339 &      16.4$\pm$3.3  &       8.1$\pm$2.4  &         42 &        105 &      12.5$\pm$3.4  &       5.4$\pm$3.1  \\

   110205A &     444643 &       2.22 &     338.4$\pm$34.2  &     105.7$\pm$7.6  &         43 &        109 &     237.3$\pm$19.7  &      94.0$\pm$9.1  \\

   110213A &     445414 &       1.46 &      37.0$\pm$6.2  &       8.7$\pm$2.0  &         57 &        142 &      28.6$\pm$4.4  &       7.6$\pm$11.6  \\

   110422A &     451901 &       1.77 &      28.1$\pm$1.4  &      11.4$\pm$0.2  &         51 &        126 &      23.8$\pm$0.8  &       9.9$\pm$0.3  \\

   110503A &     452685 &      1.613 &       9.9$\pm$3.3  &       2.9$\pm$0.3  &         54 &        134 &       7.0$\pm$1.1  &       2.0$\pm$0.3  \\

   110715A &     457330 &       0.82 &      13.0$\pm$4.6  &       1.6$\pm$0.1  &         77 &        192 &       4.0$\pm$9.4  &       1.3$\pm$0.1  \\

   110731A &     458448 &       2.83 &      42.4$\pm$12.8  &       4.9$\pm$0.2  &         37 &         91 &      31.9$\pm$19.6  &       4.7$\pm$0.2  \\

   110801A &     458521 &      1.858 &     390.8$\pm$9.1  &     333.2$\pm$5.4  &         49 &        122 &     388.1$\pm$15.1  &     307.6$\pm$175.1  \\

   110818A &     500914 &       3.36 &     101.1$\pm$17.7  &      38.0$\pm$5.3  &         32 &         80 &     100.1$\pm$23.6  &      37.2$\pm$6.8  \\

   111008A &     505054 &     4.9898 &      67.8$\pm$6.5  &      29.2$\pm$1.8  &         23 &         58 &      64.1$\pm$14.9  &      28.8$\pm$2.0  \\

   111107A &     507185 &      2.893 &      31.0$\pm$7.9  &      11.1$\pm$2.9  &         36 &         90 &      31.4$\pm$11.5  &       9.2$\pm$5.3  \\

   111228A &     510649 &      0.716 &     100.6$\pm$5.6  &      46.6$\pm$0.8  &         82 &        204 &      48.1$\pm$29.1  &       4.7$\pm$37.3  \\

   120119A &     512035 &      1.728 &     250.8$\pm$21.9  &      25.1$\pm$1.1  &         51 &        128 &     266.0$\pm$20.9  &      26.5$\pm$2.1  \\

   120326A &     518626 &      1.798 &      70.9$\pm$9.7  &       5.2$\pm$0.4  &         50 &        125 &      13.5$\pm$4.9  &       3.8$\pm$0.5  \\

   120327A &     518731 &       2.81 &      65.3$\pm$6.4  &      23.0$\pm$3.4  &         37 &         92 &      85.4$\pm$35.7  &      23.0$\pm$6.2  \\

   \hline
   090429B &   350854  &       $\sim$9.4 &   5.6$\pm$1.0  &  2.5$\pm$0.3   &  13 &  34  & 4.5$\pm$0.5   & 2.4$\pm$0.5 \\
   120521C &   522656  &       $\sim$6.0&   32.7$\pm$8.1  &  11.5$\pm$2.3  &  20 &  50  & 22.9$\pm$5.6  & 10.1$\pm$2.4 \\
   120923A &   534402  &       $\sim$8.5&   27.5$\pm$6.6  &  16.4$\pm$6.0  &  15 &  37  & 11.2$\pm$1.5  & 7.6$\pm$2.2  \\

\enddata
$^{*}$ Available in the electronic version only.\\
$^{\star}$ $E_{1}$ and $E_{2}$ correspond to the lower ($140/(1+z)$) and upper ($350/(1+z)$) limit of energy range used to measure durations $T_{90}$ and $T_{50}$.\\
\end{deluxetable}

\end{document}